\documentclass[conference,a4paper]{IEEEtran}
\usepackage{enumitem}
\usepackage{amsmath,amsthm,amssymb}
\usepackage{graphicx}
\usepackage{mathtools}
\usepackage{MnSymbol} 
\usepackage{tabularx}
\usepackage{blkarray}
\newcolumntype{H}{>{\setbox0=\hbox\bgroup}c<{\egroup}@{}}
\usepackage{blkarray}
\usepackage{subcaption}
\usepackage{tikz}
\usepackage{multirow}
\usetikzlibrary{arrows.meta}
\usetikzlibrary{decorations.pathreplacing}
\usepackage[ruled,vlined,linesnumbered]{algorithm2e}

\newcounter{example}

\usetikzlibrary{arrows}
\SetKw{KwBy}{by}
\SetKw{KwTo}{to}

\begin{document}
        
        
        \title{Polar subcodes for MIMO systems}
         \author{\IEEEauthorblockN{Liudmila Karakchieva, Peter Trifonov} 
 \IEEEauthorblockA{ITMO University, Russia\\Email: \{lvkarakchieva, pvtrifonov\}@itmo.ru}}
        \markboth{}{}

        \maketitle
        
        \begin{abstract}
       Polar-coded multiple-input multiple-output systems are investigated.
       An advanced receiver implementing joint list decoding of polar codes  and 
        QR- and MMSE-based  detectors is proposed. The approximate and exact path metrics are derived for joint list decoder of polar codes. A construction of polar subcodes for MIMO systems with cross-antenna dynamic freezing constraints is proposed. The obtained polar subcodes provide  significant performance gain compared to LDPC-coded MIMO systems with the same rate allocation. 
         \end{abstract}
        
        \section{Introduction}
Multiple-input multiple-output (MIMO) technique is used in wireless communication systems to enhance the capacity gain and reliability. It is well known that channel encoding enables to further improve the performance sacrificing the computational complexity. Therefore, error correction codes with simple construction, encoding and decoding algorithms are the most desirable to use in coded MIMO systems.

Polar codes proposed by Ar\i kan in \cite{arikan2009channel} are based on channel polarization phenomenon and achieve the capacity of binary discrete memoryless channels (BDMCs). Their simple structure and promising performance under successive cancellation list decoding \cite{niu2012crcaided, tal2015list} makes them good candidates for utilisation in multiple antenna communication.
Furthermore, polar codes are capable of providing soft output  using either optimal \cite{karakchieva2024recursive} or suboptimal techniques \cite{yuan2024near,fayyaz2014low}, which allows them to be used in iterative receivers to achieve further gains.

It is possible to implement polar-coded MIMO systems with either separate detection and decoding (SDD) scheme or joint detection and decoding (JDD) scheme.  Separate scheme \cite{dai2018polar} includes the linear receiver, where all the antennas are detected and decoded in parallel.  This is the simplest and most traditional technique, however, it demonstrates the worst performance compared to joint approach. Joint detection and decoding scheme utilizes the exchange of
information between detector and decoder \cite{jalali2020joint, yang2016joint, shen2017joint} and can be implemented either iteratively or sequentially. The first scheme uses the soft-output decoders, such as a belief propagation (BP) \cite{zhang2014simplified}, soft cancellation (SCAN) \cite{fayyaz2014low}, G-SCAN \cite{egilmez2022soft}, soft list \cite{xiang2020soft} and successive cancellation list sum-product (SCL-SP) \cite{fominykh2023efficient} algorithms, to  refine the estimate of the transmitted vector iteratively. A parallel detector, e.g. minimum mean square error detector with parallel interference cancellation (MMSE-PIC) \cite{wang1999iterative,cheng2022efficient}, is acceptable in this case. By contrast, the successive joint scheme combines successive interference cancellation (SIC) and polar decoding \cite{dai2018polar}. 

In this paper, we consider the latter scheme.  We propose  a joint list decoding algorithm with  successive interference cancellation. The problem of this approach is that coded MIMO systems with equal code rates for each antenna leads to significant performance loss, so that the correct path is lost during the first stages of decoding and CRC is not able to improve the performance. We present a rate allocation method and construction of polar codes with  cross-antenna dynamic freezing constraints. 

The rest of the paper is organized as follows. In Section \ref{sBackground}, we present the preliminaries of system model, detection algorithms and polar (sub)codes. Section \ref{sJointList} describes the proposed advanced receiver based on a joint list decoding algorithm. The proposed code construction is presented in Section \ref{sDFS}. The numeric results are presented  in Section \ref{sNumRes}. Section \ref{sConclusion} concludes the paper.

        \section{Background}
\label{sBackground}

\subsection{System model}
We consider a coded MIMO system with $N_t$ transmit  and $N_r$ receive antennas, where $N_r \ge N_t.$ The $K$ information bits $\mathbf{u}$ are coded into $\mathbf{c}=(c_{0,0},\dots,c_{0,m-1},c_{1,0},\dots,{c_{N-1,m-1}})$ and modulated into the set of $2^m$-ary symbols $x_{i,l} \in \mathcal{O}, 0\le i < N, 1 \le l \le N_t$ and then transmitted over a MIMO channel. Here, $N$ denotes the number of  time slots and $\mathcal{O}$ represents the modulation alphabet. The average transmit power  is assumed to be normalized, i.e. $E\{\mathbf{x}_i\mathbf{x}_i^H\} = \mathbf{I}_{N_t}$ for $\mathbf{x}_i = (x_{i,1},\dots, x_{i,N_t})^T.$ The encoder for $l$-th transmit antenna utilizes a $(K_l, mN)$ code, so that $\sum_{l = 1}^{N_t}K_l = K$ and the total code length is $N_tmN$. The output of MIMO channel for $i$-th time slot can be represented by 
\begin{align}\label{fMIMOOutput}\mathbf{y}_i = \mathbf{H}_i\mathbf{x}_i + \nu_i, \end{align} where $\nu_{i,j} \sim \mathcal{CN}(0,N_0), 1 \le j \le N_r ,0 \le i < N,$ and $\mathbf{H}_i$ is a $N_r \times N_t$ channel matrix of fading gains. We assume that the elements of $\mathbf{H}_i$ are known by the receiver. 
\subsection{Detection algorithms}
In this paper, we consider MIMO systems with MMSE-SIC detector \cite{vucetic2003space} and V-BLAST detector based on QR decomposition  \cite{jiang2005uniform} . Since the channel matrices for time slots vary, the reverse order for antenna processing is assigned. For the sake of simplicity, we omit the index $i$ in $\mathbf{H}_i$ within this section and consider only one time slot. 
\subsubsection{QR-based V-BLAST detector} 
Let us decompose the channel matrix $\mathbf{H}$ as \begin{equation*}\mathbf{H} = \mathbf{QR},\end{equation*} where $\mathbf{Q}$ is an $N_r \times N_t$ unitary matrix and $\mathbf{R}$ is an $N_t \times N_t$ upper triangular matrix with zero elements under the main diagonal. Taking into account such a decomposition, the received vector $\mathbf{y}$ in \eqref{fMIMOOutput} can be modified as follows
\begin{align}\label{fQR}\tilde{\mathbf{y}} =& \mathbf{Q}^H\mathbf{y}=\mathbf{Q}^H(\mathbf{H}\mathbf{x} + \mathbf{\nu})\nonumber \\=& \mathbf{R}\mathbf{x} + \mathbf{Q}^H\mathbf{\nu} = \mathbf{R}\mathbf{x} +\tilde{\mathbf{\nu}}, \tilde{\nu}_{j} \sim \mathcal{CN}(0,N_0).\end{align}
The $j$-th element of $N_t$-component column matrix $\tilde{\mathbf{y}}$  is expressed by
\begin{align*}\tilde{y}_j = R_{j,j}x_j+ \sum_{j'=j+1}^{N_t}R_{j,j'}x_{j'} +\tilde{\nu}_j.\end{align*} Consequently, the estimate of the transmitted symbol $x_j$ is given by \begin{align*}\hat{x}_j = q\left(\tilde{y}_j-\sum_{j'=j+1}^{N_t}R_{j,j'}\hat{x}_{j'}\right).\end{align*} For uncoded MIMO systems the value of  function $q(y)$ represents the discretization of $y$. In this case, the detection complexity reduces to the complexity of QR decomposition of matrix $\mathbf{H},$ that is $O(N_r^3).$

 For coded MIMO systems, we need to decompose all channel matrices $\mathbf{H}_i$ and compute $\tilde{y}_{i,j}-\sum_{j'=j+1}^{N_t}R_{i,j,j'}\hat{x}_{i,j'}$ for each time slot $0 \le i < N,$ so that the function $q(\cdot)$ requires a vector as input and its value represents the output of the decoder of the code used to encode the data. It results in overall detection complexity $O(NN_r^3).$


\subsubsection{MMSE-SIC detector} Consider the problem of minimizing  the expected value of the mean square error between the
transmitted vector $\mathbf{x}$ and its linear estimate $\hat{\mathbf{x}}=\mathbf{w}^H\mathbf{y}$ obtained from the received vector $\mathbf{y}$
\begin{align*}\min E\{||\mathbf{x}-\mathbf{w}^H\mathbf{y}||^2\},\end{align*} where
\begin{align*}\mathbf{w}^H = [\mathbf{H}^H\mathbf{H} + \frac{1}{\text{SNR}}\mathbf{I}_{Nt}]^{-1}\mathbf{H}^H\end{align*} is a $N_t \times N_r$ matrix of linear combination coefficients. The estimate of the transmitted symbols $x_j, 1 \le j \le N_t$ is obtained by
\begin{equation*}
\hat{x}_j = q\left(\mathbf{w}_j^H\mathbf{y}^j\right),
\end{equation*} where $\mathbf{y}^j = \mathbf{y} - \sum_{j' = j+1}^{N_t}\mathbf{h}_{j'}\hat{x}_{j'}, \mathbf{h}_{j'}$ is a ${j'}$-th column of $\mathbf{H}$ and $\mathbf{w}_j^H$ is a $j$-th row of $\mathbf{w}^H.$ 
 The complexity of this detection method reduces to the computation of matrix $\mathbf{w}^H$ for each transmit antenna, that is $O(N_r^3N_t)$ for uncoded MIMO system and $O(N_r^3N_tN)$ for coded MIMO system.    

\subsection{Polar (sub)codes}
\label{sSubcodes}
Polar code of length $n = 2^m$ and dimension $k$  is a linear block code being the set of codewords $\mathbf{c} = \mathbf{u}\mathbf{G},$ where $\mathbf{G} = \mathbf{F}^{\otimes m}$ is a polarizing transformation matrix, $\mathbf{F} = \begin{pmatrix}1&0\\1&1\end{pmatrix}$ and $u_j$ are predefined for some $j$ in the set of frozen indices $\mathcal{F},|\mathcal{F}| = n-k.$ This frozen set corresponds to the  most unreliable subchannels, while other subchannels are used for transmission of valuable data. The classical polar codes have $u_j = 0, j \in \mathcal{F}$ (static frozen symbols). In \cite{trifonov2016polar}, it was shown that using dynamic frozen symbols (DFS) enables one to obtain codes with higher minimum distance than classical polar codes. These codes are referred to as polar subcodes and have $u_j = \sum_{z < j}V_{s_j,z}u_z, j \in \mathcal{F}$ with the constraint matrix $V,$ where $s_j$
is the index of the row ending in column $j,$ i.e. $V_{s_j,j} = 1$ and $V_{s_j,k} = 0$ for $j < k < n.$ In \cite{trifonov2017randomized}, it was  proposed to use a randomized method to construct polar subcodes with $N_{\text{dfsA}} + N_{\text{dfsB}}$ dynamic frozen symbols of two types: dynamic frozen symbols of type B correspond to $N_{\text{dfsB}}$ most reliable bit subchannels among $n - k - N_{\text{dfsA}}$ least reliable bit subchannels $\mathcal{F}$; dynamic frozen symbols of type A correspond to $N_{\text{dfsA}}$ most reliable bit subchannels in $\mathcal{N} = [n]\backslash \mathcal{F}$ with indices of weight $w, w + 1,...,$ where $w$ is the minimal weight of integers in the remaining set of indices $\mathcal{N}.$ In this paper we assume that channel reliability of subchannels is determined by the error probability evaluated by Monte-Carlo simulations.

The successive cancellation (SC) decoding algorithm of polar codes computes subchannel probabilities 
$W_m^{(i)}(y_0^{n-1}, \hat{u}_0^{i-1}|u_i)$ for each phase $i$ and makes decisions
\begin{equation*}\hat{u}_i = \begin{cases}\arg\max_{u_i} W_m^{(i)}(y_0^{n-1}, \hat{u}_0^{i-1}|u_i), & i \nin \mathcal{F} \\ \sum_{z < i}V_{s_i,z}\hat{u}_z, & \text{otherwise}.\end{cases}\end{equation*}List decoding algorithm \cite{tal2015list} can be used to improve the performance of polar codes. In this case, the decoder keeps $L$ partial vectors of information symbols $\hat{u}_0^{i-1}$ and its scores $\mathcal{R}(\hat{u}_0^{i-1},y_0^{n-1})$. The min-sum  decoder of polar codes \cite{trifonov2018score} computes 
$$\mathcal{R}(\hat{u}_0^i,y_0^{n-1}) = \mathcal{R}(\hat{u}_0^{i-1},y_0^{n-1}) + \tau(\hat{u}_i, S_m^{(i)}(\hat{u}_0^{i-1},y_0^{n-1})),$$ where \begin{equation*}\tau(u,S) = \begin{cases}0, & \text{if $(-1)^u =$ sgn$(S)$}\\ -|S|, & \text{otherwise}\end{cases}\end{equation*} is the penalty function, and \begin{align*}S_{\lambda}^{(2i)}(\hat{u}_0^{2i-1},y_0^{n-1}) &= \text{sgn}(a)\text{sgn}(b)\min(|a|,|b|)\\S_{\lambda}^{(2i+1)}(\hat{u}_0^{2i},y_0^{n-1})&=(-1)^{\hat{u}_{2i}}a+b\end{align*} with $a = S_{\lambda-1}^{(i)}(\hat{u}_{0,e}^{2i-1}\oplus \hat{u}_{0,o}^{2i-1}, y_0^{n/2-1}), b=S_{\lambda-1}^{(i)}(\hat{u}_{0,o}^{2i-1},y_{n/2}^{n-1}),$ and $S_0^{(0)}(y) = \log\frac{W(y|0)}{W(y|1)}.$

        \tikzset{radiation/.style={{decorate,decoration={expanding waves,angle=90,segment length=4pt}}}}
\tikzstyle{input} = [coordinate]
\tikzstyle{block} = [draw, fill=white, rectangle, 
    minimum height=3em, minimum width=6em]
\tikzstyle{block2} = [draw, fill=white, rectangle, 
    minimum height=1.5em, minimum width=6em]
    \def\antenna{%
    -- +(0mm,-4.0mm) -- +(2.625mm,-7.5mm) -- +(-2.625mm,-7.5mm) -- +(0mm,-4.0mm)
}
 \def\antennaa{%
    -- +(0mm,4.0mm) -- +(2.625mm,7.5mm) -- +(-2.625mm,7.5mm) -- +(0mm,4.0mm)
}

\section{Joint List Decoder}
\label{sJointList}
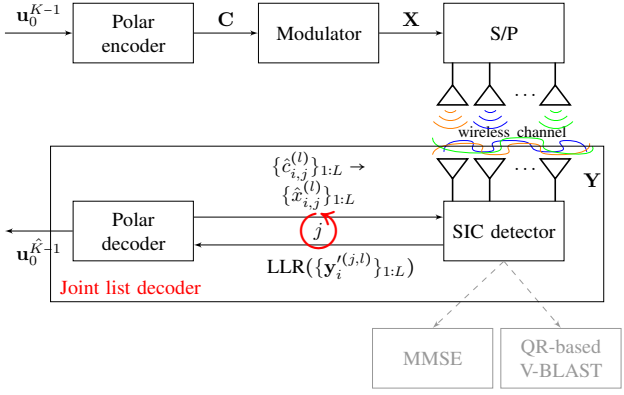
\begin{figure}[t!]
\caption{MIMO system with joint list decoder}
\scalebox{0.75}{
\begin{tikzpicture}[auto, node distance=2.25cm,text width=1.8cm, align=center,>=latex']
 \node [input, name=input] {};
 \node [block, right of=input] (encoder) {Polar encoder};
  \node [block, right of=encoder, xshift = 1cm, draw=none] (interleaver) {};
    \node [block, right of=encoder, xshift = 1cm] (modulator) {Modulator};
  \node [block, right of=modulator, xshift = 1cm] (mapper) {S/P};
  \draw [draw,->] (input) -- node {$\mathbf{u}_0^{K-1}$} (encoder);
  \draw [draw,->] (encoder) -- node {$\mathbf{C}$} (modulator);
    \draw [draw,->] (modulator) -- node {$\mathbf{X}$} (mapper);
\draw[color=black,thick] (7.85,-0.52) \antenna;
\draw[color=black,thick] (8.5,-0.52) \antenna;
\draw[color=black,thick] (9.65,-0.52) \antenna;
\node[] at (9.15,-1.1){$\dots$};
 \node [input, name=input2] at (0,-3.5){};
 \node [block, right of=input2] (decoder) {Polar decoder};
   \node [block, below of=mapper, yshift = -1.25cm] (detector) {SIC detector};
   \begin{scope}[transform canvas={yshift=.7em}]
         \draw [draw,->] (decoder.east)  --  node {$\{\hat{c}_{i,j}^{(l)}\}_{1:L} \to \{\hat{x}_{i,j}^{(l)}\}_{1:L}$} (detector.west);
\end{scope}
   \begin{scope}[transform canvas={yshift=-.7em}]
         \draw [draw,->] (detector.west)  -- node {LLR$(\{\mathbf{y}_i'^{(j,l)}\}_{1:L})$} (decoder.east);
\end{scope}
     \draw [draw,->] (decoder) -- node {$\hat{\mathbf{u}_0^{K-1}}$} (input2);
\draw[color=black,thick] (7.85,-2.975)  \antennaa;
\draw[color=black,thick] (8.5,-2.975) \antennaa;
\draw[color=black,thick] (9.65,-2.975) \antennaa;
\node[] at (9.15,-2.4){$\dots$};
\node[] at (10.3,-2.6){$\mathbf{Y}$};
\node[] at (10.2,-1.0){};
\node[text width = 5cm] at (8.9,-1.75){\footnotesize{wireless channel}};
\draw[green,radiation,decoration={angle=45}] ([xshift=-0.15cm,yshift=1.7cm]detector.north east) -- +(-90:0.5);
\draw[blue,radiation,decoration={angle=45}] ([xshift=-1.3cm,yshift=1.7cm]detector.north east) -- +(-90:0.5);
\draw[orange,radiation,decoration={angle=45}] ([xshift=-2.0cm,yshift=1.7cm]detector.north east) -- +(-90:0.5);
\draw [blue, xshift=4cm] plot [smooth, tension=2] coordinates { (3.7,-2.1) (4.05,-1.9) (4.4,-2.1) (4.85,-1.9) (5.55,-2.1) (5.85,-1.9) (6.2,-2.1)};
\draw [orange, xshift=4cm] plot [smooth, tension=2] coordinates { (3.5,-2.15)(4.1,-1.95)(4.6,-2.15)(5.1,-1.95)(5.7,-2.15)(6.25,-1.95)};
\draw [green, xshift=4cm] plot [smooth, tension=2] coordinates { (3.5,-1.85)(4.3,-2.12)(5.05,-1.85)(5.75,-2.12)(6.3,-2)};
\node[] at (5.5,-3.5){\Huge\textcolor{red}{$\circlearrowleft$}};
\node[] at (5.5,-3.5){$j$};
\node [black!40,block, below  of=detector, xshift = 1cm] (qr) {QR-based V-BLAST};
\node [black!40,block, left  of=qr] (mmse) {MMSE};
\draw [draw,->,dashed,black!40] (detector.south) -- node {} (qr.north);
\draw [draw,->,dashed,black!40] (detector.south) -- node {} (mmse.north);
\draw [draw=black] (10.5,-2) rectangle (0.8,-4.75);
\node[text width = 3cm] at (2.2,-4.5){\textcolor{red}{Joint list decoder}};
\end{tikzpicture}}
\label{fMIMOsystem}
\end{figure}

In this section, we present a joint decoding algorithm for polar-coded MIMO systems with successive interference cancellation. This method is based on successive cancellation list (SCL) decoder of polar codes with list size $L$ and utilizes a shared path metric for  all transmit antennas.  Fig. \ref{fMIMOsystem} demonstrates the block diagram for a CRC-aided MIMO system. Here, $\mathbf{u}_0^{K-1} = \{u_{p}|0 \le p \le K-1\}$ represents the vector of $K$ information bits, which are needed to distribute among the transmit antennas. More specifically, for each transmit antenna $1 \le j \le N_t$ we need to choose the frozen set providing the least  overall decoding error probability. This is a problem of polar codes construction.

Assume that $\mathbf{C}$ is a $N_t \times mN$ matrix of coded symbols, $\mathbf{X}$ is a $N_t \times N$ matrix of modulated symbols and $\mathbf{Y}$ is a matrix that represents the output of MIMO channel. The decoder obtains  the estimations $\hat{\mathbf{u}}_1^K$ for the transmitted information bits $\mathbf{u}_1^K$ starting from the $N_t$-th antenna. Let $\mathbf{y}_i'^{(j,l)}$ be the received vector for $i$-th time slot and $l$-th path in the list that was excluded from intersymbol interference provided by antennas with indices $j'>j,$ i.e. $\mathbf{y}_i'^{(j,l)} = \mathbf{y}_i - \sum_{j' =j+1}^{N_t}\mathbf{h}_{i,j'}\hat{x}_{i,j'}^{(l)},$ where $\hat{x}_{i,j'}^{(l)}$ are the decisions made earlier on the given path $l.$ The decoding process includes three steps:  
\begin{enumerate}
\item \textbf{Initialization}: $\mathbf{y}_i'^{(N_t,l_1)} \gets \mathbf{y}_i, 0\le i < mN.$ Only one path $l_1$ is active
\item \textbf{Main loop}:
for $j$ from $N_t$ to $1$
\begin{itemize}
\item If $j \neq N_t$  for each active path $l$: cancel the interference using $\hat{{x}}_{i,j-1}^{(l)}$ to update ${\mathbf{y}_i'}^{(j,l)}$ for each time slot. For QR-based V-BLAST detector we have \begin{align*}{\mathbf{y}_i'}^{(j,l)} = {\mathbf{y}_i'}^{(j-1,l)} - \mathbf{R}_{i,j-1}\hat{{x}}_{i,j-1}^{(l)},\end{align*} where  $\mathbf{R}_{i,j-1}$ is the $(j-1)$-th column of matrix $\mathbf{R}_i,$ and for MMSE-SIC detector we have \begin{align*}{\mathbf{y}_i'}^{(j,l)} = {\mathbf{y}_i'}^{(j-1,l)} - \mathbf{h}_{i,j-1}\hat{{x}}_{i,j-1}^{(l)},\end{align*} where  $\mathbf{h}_{i,j-1}$ is the $(j-1)$-th column of matrix $\mathbf{H}_i.$
\item For each active path $l$: generate LLRs for $j-$th antenna
\item Decode the information for $j$-th antenna to obtain the estimates $\{\hat{x}_{i,j}^{(l)}\}_{1:L} = \{\hat{x}_{i,j}^{(l_1)},\hat{x}_{i,j}^{(l_2)}, \dots, \hat{x}_{i,j}^{(l_L)}\}$
\end{itemize}
\item \textbf{Final decision}: Choose the path $l \in \{l_1,\dots, l_L\}$ with correct CRC 

\end{enumerate}
The  path metric for joint list decoder depends on the detection algorithm utilized in MIMO system. The derivation for two typical MIMO detectors is provided below.    
\subsection{polar-coded MIMO with QR-based V-BLAST detector}
Consider the transmission of BPSK-modulated $(m = 1)$ codewords represented by $N_t \times N$ matrix $\mathbf{X} \in \mathcal{X}$ with elements ${x}_{i,j}\in \{-\frac{1}{\sqrt{N_t}},\frac{1}{\sqrt{N_t}}\}, 1 \le j \le N_t, 0 \le i < N$ over a MIMO channel, where the output for the $i$-th time slot  is given by \eqref{fMIMOOutput}. Here, $\mathcal{X}$ describes the set of matrices having a codeword of polar code corresponding to the $j$-th antenna as the $j$-th row. 
 Let $\mathbf{H}_i=\mathbf{Q}_i\mathbf{R}_i$ be QR decompositions of channel matrices.   Taking into account \eqref{fQR}, the maximum likelihood decoding problem can be formulated as 
$$\hat{\mathbf{X}}=\arg \min_{\mathbf{X}\in \mathcal{X}}\sum_{i=0}^{N-1}||\mathbf{H}_i\mathbf{x}_i-\mathbf{y}_i||^2=\arg\min_{\mathbf{X}\in \mathcal{X}}\sum_{i=0}^{N-1}||\mathbf{R}_i\mathbf{x}_i-\tilde{\mathbf{y}}_i||^2.$$
This expression can be further transformed as follows:
\begin{align*}
\hat{\mathbf{X}}&=\arg\min_{\mathbf{X}\in \mathcal{X}}\sum_{i=0}^{N-1}\sum_{j=1}^{N_t}\Bigl(\sum_{t=j}^{N_t}R_{i,j,t}x_{i,t}-\tilde{y}_{i,j}\Bigr)^2\\&=\arg\min_{\mathbf{X}\in \mathcal{X}}\sum_{i=0}^{N-1}\sum_{j=1}^{N_t}\Bigl(R_{i,j,j}x_{i,j}+\bigl(\sum_{t=j+1}^{N_t}R_{i,j,t}x_{i,t}-\tilde{y}_{i,j}\bigr)\Bigr)^2\\&=
\arg\min_{\mathbf{X}\in \mathcal{X}}\sum_{i=0}^{N-1}\sum_{j=1}^{N_t}\bigl(R_{i,j,j}^2x_{i,j}^2+2R_{i,j,j}x_{i,j}\bigl(\sum_{t=j+1}^{N_t}R_{i,j,t}x_{i,t}-\tilde{y}_{i,j}\bigr)\\&+\bigl(\sum_{t=j+1}^{N_t}R_{i,j,t}x_{i,t}-\tilde{y}_{i,j}\bigr)^2\Bigr)\\&=
\arg\max_{\mathbf{X}\in \mathcal{X}}\sum_{j=1}^{N_t}\sum_{i=0}^{N-1}(x_{i,j}R_{i,j,j}w(\tilde{y}_{i,j},x_{i,j+1},\dots,x_{i,N_t})\\&-\frac{1}{2}w(\tilde{y}_{i,j},x_{i,j+1},\dots,x_{i,N_t})
^2),\end{align*}
where $$w(\tilde{y}_{i,j},x_{i,j+1},\dots,x_{i,N_t})=\tilde{y}_{i,j}-\sum_{t=j+1}^{N_t}R_{i,j,t}x_{i,t}$$
Let $\tilde x_{i,j}$ be hard decision values obtained from the sign of  $R_{i,j,j}w(\tilde{y}_{i,j},x_{i,j+1},\dots,x_{i,N_t})$. Then, one obtains 
\begin{align*}
\hat{\mathbf{X}}=\arg\max_{\mathbf{X}\in \mathcal{X}}\sum_{j=1}^{N_t}&\Bigl(\underbrace{\sum_{i=0}^{N-1}(x_{i,j}-\tilde x_{i,j})R_{i,j,j}w(\tilde{y}_{i,j},x_{i,j+1},\dots,x_{i,N_t})}_{E_{j, \text{V-BLAST}}(\mathbf{X})}\\&+\underbrace{\sum_{i=0}^{N-1}\tilde x_{i,j}R_{i,j,j}w(\tilde{y}_{i,j},x_{i,j+1},\dots,x_{i,N_t})}_{A_{j, \text{V-BLAST}}(\mathbf{X})}\\&-\underbrace{\sum_{i=0}^{N-1}\frac{1}{2}w(\tilde{y}_{i,j},x_{i,j+1},\dots,x_{i,N_t})
^2}_{B_{j, \text{V-BLAST}}(\mathbf{X})}\Bigr).
\end{align*}
The term $$E_{j, \text{V-BLAST}}(\mathbf{X})=-\sum_{\substack{i=0\\x_{ij}\neq \tilde x_{ij}}}^{N-1}\bigl|R_{i,j,j}w(\tilde{y}_{i,j},x_{i,j+1},\dots,x_{i,N_t})\bigr|$$
can be recognized as the correlation discrepancy of $(x_{0,j},\dots,x_{N-1,j})$ with respect to the vector $(R_{i,j,j}w(\tilde{y}_{i,j},x_{i,j+1},\dots,x_{i,N_t})|0\leq i< N)$. This is exactly the path score of the min-sum version of the Tal-Vardy list decoder. The remaining terms constitute the adjustment needed to take into account  contribution of previously processed antennae. Finally, the overall path metric for the set of codewords $\mathbf{X}$ is \begin{equation}\label{fPathMetric}P(\mathbf{X}) =\sum_{j=1}^{N_t}\bigl(E_{j, \text{V-BLAST}}(\mathbf{X}) + A_{j, \text{V-BLAST}}(\mathbf{X}) - B_{j, \text{V-BLAST}}(\mathbf{X})\bigr).\end{equation} We propose to approximate \eqref{fPathMetric} and use
\begin{equation*}
P(\mathbf{X}) \approx\sum_{j=1}^{N_t}\bigl(E_{j, \text{V-BLAST}}(\mathbf{X})\bigr).
\end{equation*}
\subsection{PC-MIMO with MMSE-SIC detector}
Assume that $\hat{\mathbf{x}}(j) \in \mathcal{X}_j$ is an estimate of modulated codeword transmitted by $j$-th antenna, that is calculated as
\begin{equation}
\hat{\mathbf{x}}(j) = \arg\min_{\mathbf{x} \in \mathcal{X}_j}\sum_{i=0}^{N-1}||\mathbf{h}_{i,j}x_{i,j} - \mathbf{y}_i^j||^2,
\end{equation} where $\mathbf{h}_{i,j}$ is a $j$-th column of matrix $\mathbf{H}_i$ and $$\mathbf{y}_i^j = \mathbf{y}_i - \sum_{j' = j+1}^{N_t}\mathbf{h}_{i,j'}\hat{x}_{i,j'} =\mathbf{h}_{i,j}x_{i,j} + \tilde{\mathbf{\eta}}_i^j.$$ Here, $ \tilde{\mathbf{\eta}}_i^j$ corresponds to the remaining noise-plus-interference (NPI). The solution of MMSE problem for antenna $j$ at time slot $i$ is obtained as   
\begin{equation}\label{eqMMSEproblem}\mathbf{w}_{i,j}^H\mathbf{y}_i^j = \mathbf{w}_{i,j}^H\mathbf{h}_{i,j}x_{i,j} +\mathbf{w}_{i,j}^H\tilde{\mathbf{\eta}}_i^j.\end{equation} Assume that $\mathcal{X}$ is the set of $N_t \times N$ matrices having the modulated codeword $\mathbf{x}(j) \in \mathcal{X}_j$ as a $j$-th row.  The maximum likelihood decoding problem for joint decoder with MMSE detector can be formulated as \begin{align}\label{fMMSEML}\hat{\mathbf{X}} &= \arg\min_{\mathbf{X} \in \mathcal{X}} \sum_{j=1}^{N_t}\sum_{i=0}^{N-1}(\mathbf{w}_{i,j}^H\mathbf{y}_i^j - \mathbf{w}_{i,j}^H\mathbf{h}_{i,j}x_{i,j})^2 \nonumber\\&= \sum_{j=1}^{N_t}\arg\min_{\mathbf{x}(j) \in \mathcal{X}_j}\sum_{i=0}^{N-1}(\mathbf{w}_{i,j}^H\mathbf{y}_i^j - \mathbf{w}_{i,j}^H\mathbf{h}_{i,j}x_{i,j})^2\nonumber\\&=\sum_{j=1}^{N_t}\arg\min_{\mathbf{x}(j) \in \mathcal{X}_j} \sum_{i=0}^{N-1}\Bigl((\mathbf{w}_{i,j}^H\mathbf{y}_i^j)^ 2 - 2\mathbf{w}_{i,j}^H\mathbf{y}_i^j\mathbf{w}_{i,j}^H\mathbf{h}_{i,j}x_{i,j} \nonumber\\&+ (\mathbf{w}_{i,j}^H\mathbf{h}_{i,j}x_{i,j})^2\Bigr)=  \sum_{j=1}^{N_t}\arg\max_{\mathbf{x}(j) \in \mathcal{X}_j} \Bigl(\sum_{i=0}^{N-1} \mathbf{w}_{i,j}^H\mathbf{y}_i^j\mathbf{w}_{i,j}^H\mathbf{h}_{i,j}x_{i,j} \nonumber\\&- \sum_{i=0}^{N-1} \frac{1}{2}(\mathbf{w}_{i,j}^H\mathbf{y}_i^j)^ 2\Bigr).\end{align} Let $\tilde x_{i,j}$ be hard decision values obtained from the sign of  $\mathbf{w}_{i,j}^H\mathbf{y}_i^j\mathbf{w}_{i,j}^H\mathbf{h}_{i,j}. $ Therefore, we can rewrite \eqref{fMMSEML} as
\begin{align*}  \hat{\mathbf{X}} &= \sum_{j=1}^{N_t}\arg\max_{\mathbf{x}(j) \in \mathcal{X}_j}  \Bigl(\underbrace{\sum_{i=0}^{N-1}(x_{i,j}- \tilde{x}_{i,j})\mathbf{w}_{i,j}^H\mathbf{y}_i^j\mathbf{w}_{i,j}^H\mathbf{h}_{i,j}}_{E_{j,\text{MMSE}}(\mathbf{X})} \\&+ \underbrace{\sum_{i=0}^{N-1}\tilde{x}_{i,j}\mathbf{w}_{i,j}^H\mathbf{y}_i^j\mathbf{w}_{i,j}^H\mathbf{h}_{i,j}}_{A_{j,\text{MMSE}}(\mathbf{X})} - \underbrace{\sum_{i=0}^{N-1} \frac{1}{2}(\mathbf{w}_{i,j}^H\mathbf{y}_i^j)^ 2}_{B_{j,\text{MMSE}}(\mathbf{X})}\Bigr).\end{align*}
The term 
$$E_{j,\text{MMSE}}(\mathbf{X})=-\sum_{\substack{i=0\\x_{ij}\neq \tilde x_{ij}}}^{N-1}\bigl|\mathbf{w}_{i,j}^H\mathbf{y}_i^j\mathbf{w}_{i,j}^H\mathbf{h}_{i,j}\bigr|$$ represents the correlation discrepancy of $(x_{0,j},\dots,x_{N-1,j})$ with respect to the vector $(\mathbf{w}_{i,j}^H\mathbf{y}_i^j\mathbf{w}_{i,j}^H\mathbf{h}_{i,j}|0\leq i < N).$ Eventually, we obtain the overall path metric 
$$P_{\text{MMSE}}(\mathbf{X}) = \sum_{j = 1}^{N_t}(E_{j,\text{MMSE}}(\mathbf{X})+A_{j,\text{MMSE}}(\mathbf{X})-B_{j,\text{MMSE}}(\mathbf{X}))$$ and its approximation
$$P_{\text{MMSE}}(\mathbf{X}) \approx \sum_{j = 1}^{N_t}E_{j,\text{MMSE}}(\mathbf{X}).$$

Simulation results show that using approximate path scores for MIMO systems with both QR-based V-BLAST and MMSE detectors results in negligible performance loss.

        \section{Construction of Polar Subcodes}
\label{sDFS}
In this section, we present the code construction algorithm for  MIMO systems with $N_{\text{dfsA}} + N_{\text{dfsB}}$ cross-antenna dynamic frozen symbols. Taking into account the notation in Section \ref{sSubcodes},  assume that 
DFS of type B correspond to $N_{\text{dfsB}}$ most reliable bit subchannels among $N_tmN - K - N_{\text{dfsA}}$ least reliable bit subchannels $\mathcal{F}$ and $\mathcal{N} = [N_tmN]\backslash \mathcal{F}.$
Assume that the joint list decoder consider polar subcodes of length $mN$ for each antenna $1 \le j \le N_t$ as a long polar subcode of length $N_tmN$. For MIMO system the weights of indices $w_i, 0 \le i < N_tmN$ for each antenna are the same, i.e. $w_k = w_{k + mN} =...= w_{k+(N_t-1)mN}$ with $w_i = \mathbf{wt}(i \text { mod } mN).$  
While assigning the dynamic frozen symbols of type A, the selecting order of subchannels corresponding to the indices of the same weight $\mathbf{w}$ is strictly depends on the order of antenna processing. Since the order of antenna processing is reverse, one needs to start with the first antenna and select all indices of weight $\mathbf{w}$  starting with the largest integers. When all indices of weight $\mathbf{w}$ for the first antenna are assigned, we can go to the second antenna (see Algorithm \ref{AlgDFSA}).

As the result, we obtain the following algorithm for construction of polar subcodes for MIMO systems.

\begin{algorithm}[t!] 
 \caption{SetDFS-A$(N_t, m, N, \mathcal{N}, N_{\text{dfsA}})$}
          \label{AlgDFSA}  
\For{$i \gets 0$ \KwTo $N_tmN-1$  \KwBy $1$}
{$w_i \gets  \mathbf{wt}(i \text { mod } mN)$;}
$w \gets \min_{i \in \mathcal{N}}w_i$\;
$\mathbf{w} \gets w$\;
$p \gets 0$\;
\While{$(p < N_{\text{dfsA}})$}{
\For{$j \gets N_t$ \KwTo $1$ \KwBy $-1$}
{\For{$k \gets jmN$ \KwTo $(j-1)mN + 1$ \KwBy $-1$}
{\If{$(w_k = \mathbf{w} \text{ and }k \in \mathcal{N})$}{$z_p \gets k$\; $p \gets p + 1$\;\If{$p = N_{\text{dfsA}}$}{\Return $\{z_1,\dots,z_{N_{\text{dfsA}}}\}$\;}}
}}$\mathbf{w} \gets \mathbf{w} + 1$\;}
\end{algorithm} 
 
\begin{itemize}
\item Let $\mathcal{F} \subset [N_tmN]$ be the set of $N_tmN-K-N_{\text{dfsA}}$ indices of least reliable bit subchannels
\item $\mathcal{N} \gets [N_tmN]\backslash\mathcal{F}$
\item DFS-B:
\begin{itemize}
\item Let $\widehat{\mathcal{F}}$ be the set of  $N_{\text{dfsB}}$ elements of $\mathcal{F}$ corresponding to the most reliable subchannels 
\item Set static frozen symbols $\widetilde{\mathcal{F}} \gets \mathcal{F} \backslash \widehat{\mathcal{F}}$
\end{itemize}
\item DFS-A:
\begin{itemize}
\item $\mathcal{Z} \gets \text{SetDFS-A}(N_t, m,N,\mathcal{N},N_{\text{dfsA}})$ (Agorithm \ref{AlgDFSA})
\end{itemize}
\item Set DF constraints for symbols with indices in $\widehat{\mathcal{F}}, \mathcal{Z}$ taking into account that antennas are processed in reverse order and time slots for each antenna are processed in the direct order 
\end{itemize}
It must be recognized that the order of selectiong of type A dynamic frozen symbols strictly depends on the order of antenna processing.

The Algorithm \ref{AlgDFSA} is also applicable for block fading channels, where $\mathbf{H}_i = \mathbf{H}$ for all $0 \le i < N.$ In this case, one can set an optimal processing order according to post detection signal-to-noise-ratio for V-BLAST detector \cite{wolniansky1998v} or signal-to-interference-and-noise ratio for MMSE detector \cite{benjebbour2001comparison}.  

        \section{Numeric Results}
\label{sNumRes}
In this section, we investigate the frame error rate (FER) performance for a MIMO system with $N_t = N_r = 4$ and QPSK modulation. Both QR-based V-BLAST and MMSE detectors are considered. 

\begin{figure}[t!]
\begin{center}
\includegraphics[width=0.48\textwidth]{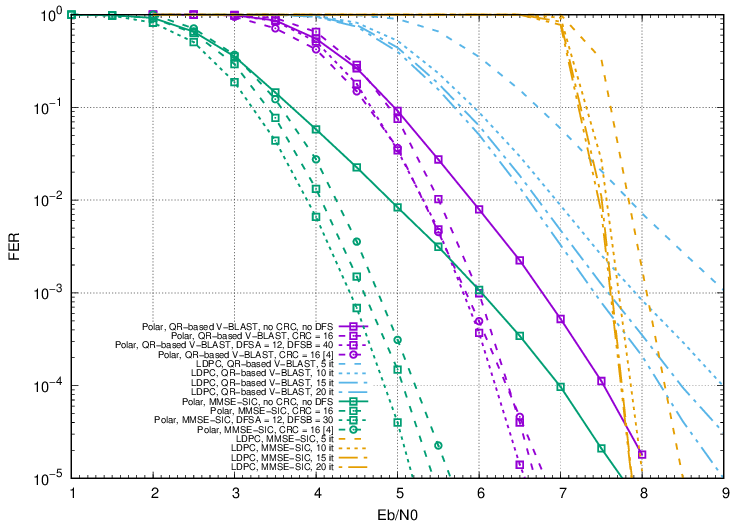}\end{center}
\caption{FER performance for polar-coded and LDPC-coded MIMO systems with $mN = 256$}
\label{fFER}
\end{figure}

Assume that $K_j$ is the dimension of polar (sub)code  for $j$-th antenna and $\mathcal{K}_{\text{D}} = \{K_1, K_2,...,K_{N_t}\}$ is the distribution of dimensions for polar-coded MIMO system obtained using the detector D. 
Fig. \ref{fFER} demonstrates FER performance for polar-coded and LDPC-coded MIMO systems with $mN = 256$, where dimension distributions for both LDPC codes and polar codes correspond to $\mathcal{K}_{\text{V-BLAST}}= \{180,154,115,63\}$ and $\mathcal{K}_{\text{MMSE}} = \{159,138,118,97\}.$ The joint list algorithm with list size $L =32$ and layered belief propagation algorithm with $I \in \{5,10,15,20\}$ iterations are used for decoding of polar codes and LDPC codes, respectively. Polar subcodes constucted by the method described in Section \ref{sDFS} and CRC-aided polar codes with $L_{\text{CRC}} = 16$ are also considered for comparison. Fig. \ref{fFER} also includes the results obtain in \cite{dai2018polar} for the same parameters. It can be seen that polar codes with dynamic frozen symbols outperform CRC-aided polar codes. The proposed polar-coded MIMO system with CRC shows better FER performance compared to the system described in \cite{dai2018polar}. Moreover, joint list decoder with MMSE-SIC detector provides significantly better performance compared to joint list decoder with QR-based V-BLAST detector. It can be also seen that polar-coded MIMO systems  outperform LDPC-coded MIMO systems with the same parameters. 
                
        \section{Conclusion}
\label{sConclusion}
In this paper, a joint list decoding and detection method for polar-coded MIMO\ systems was presented, and a construction of polar subcodes with cross-antenna dynamic frozen symbols was introduced.  

 It was shown that using approximate path metric in the proposed joint list decoding and detection algorithm  does not result in any noticeable  performance degradation. The proposed MIMO system with polar subcodes provide better performance compared to LDPC codes with the same rate allocation.

\bibliographystyle{ieeetran}

\end{document}